\documentclass[aps,pra,preprint,groupedaddress,showpacs]{revtex4-1}
\usepackage{epsfig,amsmath,graphicx,amssymb,dcolumn,bm,multirow}
\usepackage[usenames]{color}
\begin{document}

\title{Crossover from Electromagnetically Induced Transparency to Autler-Townes Splitting in Open Ladder Systems with Doppler Broadening}

\author{Chaohua Tan}
\affiliation{State Key Laboratory of Precision Spectroscopy and
Department of Physics, East China Normal University, Shanghai
200062, China}

\author{Guoxiang Huang}
\email[Email: ] {gxhuang@phy.ecnu.edu.cn} \affiliation{State Key
Laboratory of Precision Spectroscopy and Department of Physics, East
China Normal University, Shanghai 200062, China}
\date{\today}

\begin{abstract}

We propose a general theoretical scheme to investigate the crossover from electromagnetically induced transparency (EIT) to Autler-Townes splitting (ATS) in open ladder-type atomic and molecular systems with Doppler broadening. We show that when the wavenumber ratio $k_c/k_p\approx -1$, EIT, ATS, and EIT-ATS crossover exist for both ladder-I and ladder-II systems, where $k_c$ ($k_p$) is the wavenumber of control (probe) field. Furthermore, when $k_c/k_p$ is far from $-1$  EIT can occur but ATS is destroyed if the upper state of the ladder-I system is a Rydberg state. In addition, ATS exists but EIT is not possible if the control field used to couple the two lower states of the ladder-II system is a microwave field. The theoretical scheme developed here can be applied to atoms, molecules, and other systems (including Na$_2$ molecules, and Rydberg atoms), and the results obtained may have practical applications in optical information processing and transformation.


\end{abstract}
\maketitle

\section{INTRODUCTION}{\label{sec1}}


In recent years, much attention has been paid to the study of  electromagnetically induced
transparency (EIT), a quantum interference effect induced by a strong control
field, by which the optical absorption of a probe field in resonant three-level atomic systems
can be largely suppressed. In addition to the interest in fundamental research, EIT has
many important applications in slow light and quantum storage, nonlinear optics at low-light
level, precision laser spectroscopy, and so on~\cite{Fleischhauer2005}.

The most prominent character of EIT is the opening of a transparency window
in probe-field absorption spectra. However, the occurrence of transparency
window is not necessarily due to EIT effect.
In 1955, Autler and Townes~\cite{at} showed that the absorption spectrum of
molecular transition can split into two Lorentzian lines (doublet)
when one of two levels involved in the transition is coupled to a
third one by a strong microwave field. Such doublet is now called
Autler-Townes splitting (ATS) and has also been intensively investigated in
atomic and molecular spectroscopy~\cite{coh}.

Although both EIT and ATS effects can open transparency windows in probe
absorption spectra, the
physical mechanisms behind them are quite different. EIT is resulted
from a quantum {\it destructive} interference between two competing
transition pathways, whereas ATS is a dynamic Stark shift caused by
a gap between two resonances. Usually, it is not easy to distinguish EIT
and ATS by simply looking at the appearance of absorption spectra.

Because EIT and ATS are two typical phenomena appeared widely in laser spectroscopy
and have many applications, it is necessary to develop an effective technique to distinguish the
difference between them. In 1997, Agarwal~\cite{Agarwal1997}  proposed a
spectrum decomposition method, by which the probe-field absorption spectra of cold
three-level atomic systems  were decomposed into two absorptive contributions
plus two interference contributions. Recently, this method was used to clarify EIT
and ATS in a more general way~\cite{Anisimov2008,Tony2010,Anisimov2011}.
In a recent work, an experimental investigation on EIT-ATS crossover
was carried out~\cite{giner}. Very recently, the spectrum
decomposition method was adopted to investigate
the EIT-ATS crossover in $\Lambda$- and $V$-type molecular systems with Doppler
broadening~\cite{tan2013,Zhu2013}.

In the study of EIT and ATS, several typical three-level systems (i.e.
$\Lambda$, $V$, and ladder)~\cite{note00}  are widely used. For ladder systems, 
there are two typical configurations, 
with the level diagrams and excitation schemes shown in Figs.~\ref{model}(a) and \ref{model}(b) below,
called here as ladder-I (or upper-level-driven ladder system; Fig.~\ref{model}(a)\,)
and ladder-II (or lower-level-driven ladder system; Fig.~\ref{model}(b)\,), respectively.
The so-called upper-level-driven (lower-level-driven) means that the control field couples the
two upper (lower) levels of the system.
In recent years, much interest has been focused on the Rydberg excitations in cold and hot atomic gases,
where ladder-type excitation schemes are widely employed~\cite{saf,pri1,sevi,moh1,moh2,wea,rai,pri}. 

In this article, we propose a general theoretical scheme to investigate the crossover from  
electromagnetically induced transparency (EIT) to Autler-Townes splitting (ATS) 
in open ladder-type atomic and molecular systems with Doppler broadening.  
We show that when the wavenumber ratio $k_c/k_p\approx -1$, EIT, ATS, and EIT-ATS crossover exist for both ladder-I and ladder-II systems, where $k_c$ ($k_p$) is the wavenumber of control (probe) field. Furthermore, when $k_c/k_p$ is far from $-1$  EIT can occur
but ATS is destroyed if the upper state of the ladder-I system is a Rydberg state. In addition,
ATS exists but EIT is not possible if the control field used to couple the two lower states of the ladder-II system is a microwave field. The theoretical scheme developed and the results obtained here can be applied to various ladder systems (including hot gases of Rubidium atoms, Na$_2$ molecules, and Rydberg atoms).

Before proceeding, we note that many studies exist on the study of ladder systems.
Except for EIT and ATS~\cite{Agarwal1997,Tony2010,saf,moh1,moh2,wea,rai,pri,pri1,sevi,Lazoudis2008,Yang1997,Yong1995,Julio1995,
Lee2000,Jason2001,Qi2002,Ahmed2006,Ahmed2007,Ray2007,Moon,Kubler2010,gor,petro,ate}, other related
investigations have also been carried out, including Rabi oscillations~\cite{dudin,Huber2011},
coherent population transfer~\cite{schemp}, quantum nonlinear optics at single-photon
level~\cite{Peyr}, fast entanglement generation~\cite{Bari}, and microwave electrometry with Rydberg
atoms~\cite{Sedl}. However, to the best of our knowledge no systematic analysis on the crossover from EIT to ATS in ladder systems has been carried out up to now; furthermore, no theory on the EIT-ATS crossover in open ladder systems with Doppler broadening has been presented. Our theoretical scheme is valid for both atoms, molecules, and other systems, and can
elucidate various quantum interference characters (EIT, ATS, and EIT-ATS crossover)
in a clear way. The results obtained here are not only useful for understanding the detailed
feature of quantum interference in multi-level systems and guiding new experimental findings,
but also may have promising applications in atomic and molecular spectroscopy, light and quantum
information processing, etc.

The article is arranged as follows. In the next section, we describe the theoretical model. In
Sec.~\ref{sec3} and Sec.~\ref{sec4}, the quantum interference characters of the ladder-I
and ladder-II systems are analyzed, respectively. Finally, in the last section
we summarize the main results obtained in this work.

\section{MODEL}{\label{sec2}}

\begin{figure}
\includegraphics[scale=0.45]{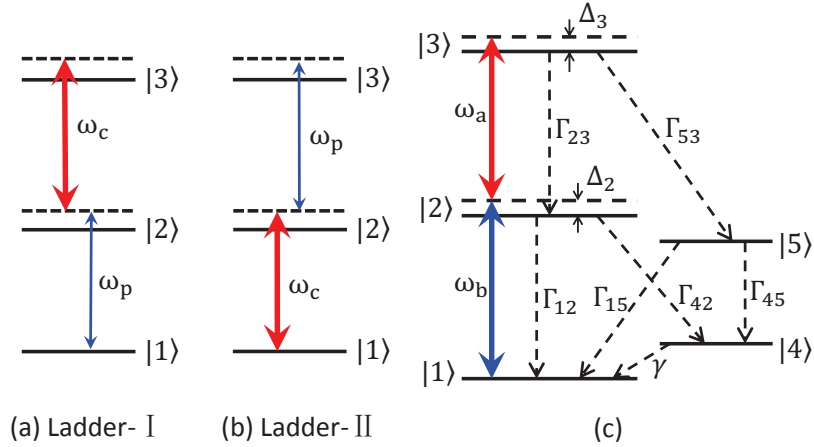}\\
\caption{\footnotesize (Color online) (a) Ladder-I system, where states
$|3\rangle$ and $|2\rangle$ are coupled by the control field
with center angular frequency $\omega_c$, and states $|2\rangle$ and
$|1\rangle$ are coupled by the probe field with center angular
frequency $\omega_p$; (b) Ladder-II system. (c) Open
ladder system. The state $|2\rangle$
couples to the state $|3\rangle$ by field $a$ (with center angular
frequency $\omega_a$) and the ground state $|1\rangle$ by field
$b$ (with center angular  frequency $\omega_b$). $\Delta_{2}$ and $\Delta_{3}$
are detunings, $\Gamma_{jl}$ are population decay rates from
$|l\rangle$ to $|j\rangle$, and $\gamma$ is the transit rate. Particles
occupying the state $|2\rangle$ ($|3\rangle$) may decay to
other states besides $|1\rangle$ ($|2\rangle$). Levels $|4\rangle$ and $|5\rangle$
denote these other states rendering the system open.}
\label{model}
\end{figure}
%

We consider a hot gas consisting of atoms or molecules, where particles have three
resonant levels (i.e. ground state $|1\rangle$,
intermediate state $|2\rangle$, and upper state  $|3\rangle$)
with a ladder configuration (Fig.~\ref{model}(c)\,)~\cite{Ahmed2007}.
Especially, the upper state $|3\rangle$ may be a Rydberg state.
Two laser fields with central angular frequency $\omega_a$ and $\omega_b$ couple
to the transition $|2\rangle\leftrightarrow|3\rangle$ and
$|1\rangle\leftrightarrow|2\rangle$, respectively. The electric field
vector is $\mathbf{E}=\sum_{l=a,b}\mathbf{e}_l{\cal
E}_l {\rm exp}[i({\bf k}_l\cdot {\bf r}-\omega_lt)]+$c.c., where
$\mathbf{e}_l$  $(k_l)$ is the unit polarization vector (wavenumber) of
the electric field component with the envelope ${\cal E}_l$ $ (l=a,b)$.

We assume the system is open, i.e. particles occupying the state $|2\rangle$ ($|3\rangle$)
can follow various relaxation  pathways and decay into other states besides $|1\rangle$  ($|2\rangle$).
For simplicity, all these other states are represented by states $|4\rangle$ and
$|5\rangle$~\cite{Ahmed2007}. In the figure, $\Delta_2$ and $\Delta_3$ are detunings, $\Gamma_{jl}$
is the population decay rate from state $|l\rangle$ to state $|j\rangle$, $\gamma$ is
the beam-transit rate added to account for the rate with which particles
escape the interaction region (significant only for the level $|4\rangle$ since
it cannot radiatively decay).

Under electric-dipole and rotating-wave
approximations, the interaction Hamiltonian of the system in interaction picture reads
\begin{equation}\label{H_EIT}
{\cal H}_{\rm
int}=-\hbar(\Omega_ae^{i[\mathbf{k}_a\cdot(\mathbf{r}+\mathbf{v}t)-\omega_at]}
|3\rangle\langle2|+\Omega_be^{i[\mathbf{k}_b\cdot(\mathbf{r}+\mathbf{v}t)
-\omega_bt]}|2\rangle\langle1|+{\rm h.c.}),
\end{equation}
where $\Omega_a=\boldsymbol{\mu}_{32}\cdot{\cal E}_a/\hbar$
($\Omega_b=\boldsymbol{\mu}_{21}\cdot{\cal E}_b/\hbar$)
is the half Rabi-frequency of the field $a$  (field $b$),
with $\boldsymbol{\mu}_{jl}$ being
the electric-dipole matrix element associated with the transition
from the state $|l\rangle$ to the state $|j\rangle$. The optical Bloch
equation in the interaction picture is
\begin{subequations} \label{dme1}
\begin{eqnarray}
&&i\frac{\partial}{\partial
t}\sigma_{11}-i\Gamma_{12}\sigma_{22}-i\gamma\sigma_{44}-i\Gamma_{15}\sigma_{55}
+\Omega_{b}^{*}\sigma_{21}-\Omega_{b}\sigma_{21}^{*}=0,\\
&&i\frac{\partial}{\partial
t}\sigma_{22}+i\Gamma_{2}\sigma_{22}-i\Gamma_{23}\sigma_{33}
+\Omega_{b}\sigma_{21}^{*}+\Omega_{a}^{*}\sigma_{32}-\Omega_{b}^{*}\sigma_{21}
-\Omega_{a}\sigma_{32}^{*}=0,\\
&&i\frac{\partial}{\partial
t}\sigma_{33}+i\Gamma_{3}\sigma_{33}+\Omega_{a}\sigma_{32}^{*}-\Omega_{a}^{*}\sigma_{32}=0,\\
&&i\frac{\partial}{\partial
t}\sigma_{44}+i\gamma\sigma_{44}-i\Gamma_{42}\sigma_{22}-i\Gamma_{45}\sigma_{55}=0,\\
&&i\frac{\partial}{\partial
t}\sigma_{55}+i\Gamma_{5}\sigma_{55}-i\Gamma_{53}\sigma_{33}=0,\\
&&(i\frac{\partial}{\partial
t}+d_{21})\sigma_{21}+\Omega_{a}^{*}\sigma_{31}+\Omega_{b}(\sigma_{11}-\sigma_{22})=0,\\
&&(i\frac{\partial}{\partial
t}+d_{31})\sigma_{31}-\Omega_{b}\sigma_{32}+\Omega_{a}\sigma_{21}=0,\\
&&(i\frac{\partial}{\partial
t}+d_{32})\sigma_{32}-\Omega_{b}^{*}\sigma_{31}+\Omega_{a}(\sigma_{22}-\sigma_{33})=0,
\end{eqnarray}
\end{subequations}
where $d_{21}=-\mathbf{k}_b\cdot\mathbf{v}+\Delta_2+i\gamma_{21}$,
$d_{31}=-(\mathbf{k}_b+\mathbf{k}_a)\cdot\mathbf{v}+\Delta_3+i\gamma_{31}$,
$d_{32}=-\mathbf{k}_a\cdot\mathbf{v}+\Delta_3-\Delta_2+i\gamma_{32}$
with $\gamma_{jl}=(\Gamma_{j}+\Gamma_{l})/2+\gamma_{jl}^{{\rm col}}\
(j,l=1,2,3)$. Here, $\bf{v}$ is the thermal velocity of the particles, 
$\Gamma_{j}$ denotes the total population decay rates
out of levels $|j\rangle$, defined by $\Gamma_{j}=\sum_{l\neq j}\Gamma_{lj}$.
The quantity $\gamma_{jl}^{{\rm col}}$ is the dephasing rate due to processes
that are not associated with population transfer, such as elastic
collisions.

The evolution of the electric field is governed by the Maxwell
equation
\begin{equation}\label{ME}
\nabla^2 {\bf E}-\frac{1}{c^2}\frac{\partial^2{\bf E}}{\partial
t^2}=\frac{1}{\epsilon_0c^2}\frac{\partial^2 {\bf P}}{\partial t^2}.
\end{equation}
Due to the Doppler effect, the electric polarization intensity of
the system reads
\begin{equation}\label{P}
{\bf P}={\cal N}\int_{-\infty}^{\infty}dv f(v)\left[
\boldsymbol{\mu}_{12}\sigma_{21} e^{i(k_b z-\omega_b
t)}+\boldsymbol{\mu}_{23}\sigma_{32} e^{i(k_a z-\omega_a t)}+{\rm
c.c.}\right],
\end{equation}
where ${\cal N}$ is particle concentration and $f(v)=e^{-(v/v_T)^2}/(\sqrt{\pi}v_T)$ 
is Maxwellian velocity distribution function, where $v_T=(2k_BT/M)^{1/2}$
is the most probable speed at temperature $T$ with
$k_B$ the Boltzmann constant and $M$ the particle mass.
For simplicity and without loss of generality, we have assumed the two laser fields
propagate along the $z$ direction with a counter-propagating configuration,
i.e. ${\bf k}_{a,b}=(0,0,k_{a,b})$ with $k_b=-k_a$ in order to suppress the first-order
Doppler effect.

Note that the model given above is valid also for a closed ladder system,
which can be obtained by simply taking $\Gamma_{15}=\Gamma_{42}=\Gamma_{45}=\Gamma_{53}=\gamma=0$;
furthermore, if the system is not only closed but also cold, one has
$\Gamma_{15}=\Gamma_{42}=\Gamma_{45}=\Gamma_{53}=\gamma=0$ and $f(v)=\delta (v)$.

\section{Quantum interference character of ladder-\uppercase\expandafter{\romannumeral1}
system}\label{sec3}

\subsection{Linear dispersion relation}\label{sec3a}

When the laser field $a$ (field $b$) is taken as the control (probe) field, the system is
the ladder-\uppercase\expandafter{\romannumeral1} system (i.e. $\omega_a=\omega_c$,
$\omega_b=\omega_p$; see Fig.~\ref{model}(a)\,). In this case, under slowly varying
envelope approximation (SVEA) the Maxwell Eq.~(\ref{ME}) is reduced to the form
\begin{equation}\label{maxwell1}
i\left(\frac{\partial}{\partial
z}+\frac{1}{c}\frac{\partial}{\partial t}\right)\Omega_b
+\kappa_{12}\int_{-\infty}^{\infty} dv f(v)\sigma_{21}(v)=0,
\end{equation}
where $\kappa_{12}={\cal
N}\omega_b|\boldsymbol{\mu}_{21}|^2/(2\hbar\varepsilon_0 c)$ with
$c$ is the light speed in vacuum.

The base state (zero-order) solution of the system, i.e. the steady-state solution of the MB
Eqs.~(\ref{dme1}) and (\ref{maxwell1}) for $\Omega_b=0$ 
is given by $\sigma_{11}^{(0)}=1$,  $\sigma_{jl}^{(0)}=0$ ($j,l\neq 1$).
When the probe field is switched on, the system will involve into time-dependent state.
At the first order of $\Omega_b$, the population and the coherence between
the states $|2\rangle$ and $|3\rangle$ are not changed, but 
\begin{subequations}\label{1order}
\begin{eqnarray}
&&\Omega_b^{(1)}=Fe^{i\theta}\\
&&\sigma_{21}^{(1)}=\frac{\omega+d_{31}}{|\Omega_a|^2-(\omega+d_{21})(\omega+d_{31})}F
e^{i\theta},\\
&&\sigma_{31}^{(1)}=-\frac{\Omega_a}{|\Omega_a|^2-(\omega+d_{21})(\omega+d_{31})}F
e^{i\theta},
\end{eqnarray}
\end{subequations}
here $F$ is a constant and $\theta=K(\omega)z-\omega t$. The linear
dispersion relation $K(\omega)$ reads
\begin{equation}\label{LDa}
K(\omega)=\frac{\omega}{c}+\kappa_{12}\int_{-\infty}^{\infty} dv
f(v)\frac{\omega+d_{31}}{|\Omega_a|^2-(\omega+d_{21})(\omega+d_{31})}.
\end{equation}
The integrand in the dispersion relation (\ref{LDa}) depends on two
factors. The first is the ac Stark effect induced by the control
field, reflected in the denominator, corresponding to the appearance
of dressed states out of states $|2\rangle$ and $|3\rangle$,  by
which two Lorentzian peaks in the probe-field absorption spectrum
are shifted from their original positions. The second is the Doppler
effect, reflected by $d_{ij}=d_{ij}(v)$ and the velocity
distribution $f(v)$, which results in an inhomogeneous broadening
in Im($K$) (the imaginary part of $K$).

\begin{figure}
\includegraphics[scale=0.45]{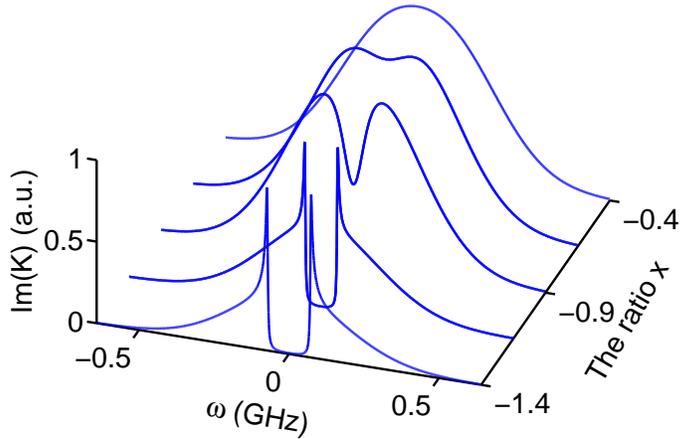}
\caption{\footnotesize (Color online) The probe-field
absorption spectrum Im($K$) of the ladder-I system as a function of $\omega$
and the wavenumber ratio $x$.} \label{EIT3D}
\end{figure}

The lineshape of Im($K$) depends strongly on the wavenumber ratio
$x=k_a/k_b$.  Fig.~\ref{EIT3D} shows
%
%
the numerical result of Im($K$) as a function of $\omega$ and $x$. 
The system parameters are chosen as $\Gamma_2=6$ MHz, $\Gamma_3=1$ MHz, $\gamma=0.5$ MHz,
$\gamma_{ij}^{\rm col}=1$ MHz, and $\Omega_a=80$ MHz.
We see that Im($K$) undergoes a transition from a deep, wide transparency window (doublet) to
a single absorption peak when $x$ changes from $-1.4$ to
$-0.4$. Since Fig.~\ref{EIT3D} is obtained by a numerical calculation, it is not easy
to get a clear and definite conclusion on the quantum interference characters of the system.
Thus we turn to an analytical approach by using the method developed
in Refs.~\cite{Agarwal1997,Anisimov2008,Tony2010,Anisimov2011,tan2013,Zhu2013}.

\subsection{EIT-ATS crossover in hot Rubidium atomic gases}

In many experimental studies on EIT or EIT-related effects in the
ladder-I system with Doppler broadening, the excitation scheme
$5S_{1/2} \rightarrow 5P_{3/2} \rightarrow 5D_{5/2}$ of $^{87}$Rb atoms was
adopted, such as did in Refs.~\cite{Julio1995,Moon}. In this situation, the wavenumber
ratio $x=-1$, and the integration in Eq.~(\ref{LDa}) can be carried out
analytically by using the residue theorem when the Maxwellian
velocity distribution is replaced by the modified Lorentzian velocity
distribution $f(v)=v_T/\left[\sqrt{\pi}(v^2+v_T^2)\right]$. Such
technique has been widely employed by many 
authors~\cite{tan2013,Zhu2013,Elena2002,Lee2003,LiHuang2010}.

\begin{figure}
\includegraphics[scale=0.35]{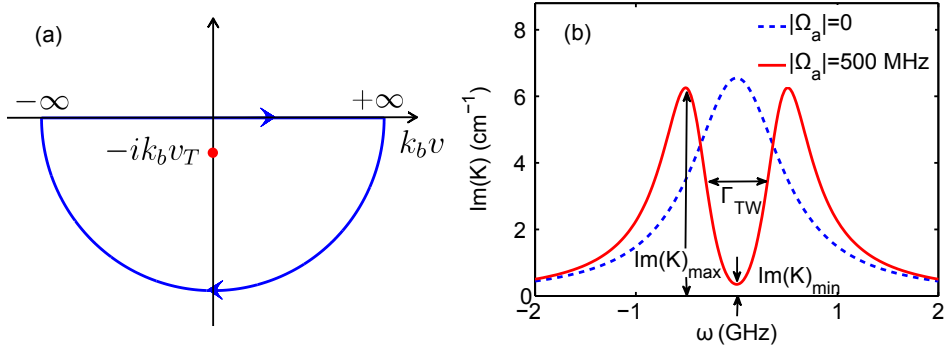}
\caption{(Color online) (a) The pole $(0,-ik_bv_T)$ of the
integrand in Eq.~(\ref{LDa}) in the lower half complex plane of $v$. The
closed curve with arrows is the contour chosen for calculating the
integration in Eq.~(\ref{LDa}) by using the residue theorem. (b)
Probe-field absorption spectrum Im($K$) as a function of $\omega$ for the hot
ladder-I system with wavenumber ratio $x=-1$. The solid (dashed) line is for $|\Omega_a|=500$ MHz
($|\Omega_a|=0$). Definitions of Im$(K)_{\rm min}$, Im$(K)_{\rm
max}$, and the width of transparency window $\Gamma_{\rm TW}$ are
indicated in the figure.}\label{fig:LM}
\end{figure}

Note that the integrand in the second term of the Eq.~(\ref{LDa}) has only one pole
$k_bv=-ik_bv_T=-i\Delta\omega_D$ in the lower half complex plane of $v$.
Considering the contour integration shown in Fig.~\ref{fig:LM}(a)
%
%
and using the residue theorem, we obtain the exact result
\begin{equation}\label{Ka1}
K(\omega)=\frac{\omega}{c}+\frac{\sqrt{\pi}\kappa_{12}(\omega
+i\gamma_{31})}{|\Omega_a|^2-(\omega+i\gamma_{21}+i\Delta\omega_D)(\omega+i\gamma_{31})},
\end{equation}
with$\Delta_2=\Delta_3=0$.
Explicit expression of $K(\omega)$ for nonvanishing $\Delta_2$ and $\Delta_3$
can also be obtained but lengthy and thus omitted here.

Fig.~\ref{fig:LM}(b) shows the profile of Im($K$) as a function of $\omega$. The
dashed (solid) line is for the case of $|\Omega_a|=0$ ($|\Omega_a|=500$ MHz) for
$\Gamma_{2}=6$ MHz, $\Gamma_{3}=1$
MHz, $\gamma=0.5$ MHz, $\gamma_{jl}^{{\rm col}}=1$ MHz,
$\Delta\omega_D=270$ MHz and $\kappa_{12}=1\times10^{9}$
cm$^{-1}$s$^{-1}$, used in Ref.~\cite{Julio1995}. We see that the probe-field absorption
spectrum for $|\Omega_a|=0$ has only a single peak. However, a transparency
window is opened for $\Omega_a=500$ MHz. The minimum (Im$(K)_{\rm min}$), maximum (Im$(K)_{\rm
max}$), and width of transparency window ($\Gamma_{\rm TW}$) are
defined in the figure.

Equation (\ref{Ka1}) can be written as the form
\begin{equation}\label{Kd1}
K(\omega)=\frac{\omega}{c}-\sqrt{\pi}\kappa_{12}\frac{\omega
+i\gamma_{31}}{(\omega-\omega_+)(\omega-\omega_-)},
\end{equation}
with
\begin{equation}
\omega_{\pm}=\frac{1}{2}\left\{-i(\gamma_{21}+\gamma_{31}+\Delta\omega_D)
\pm 2\left[ |\Omega_a|^2-|\Omega_{\rm ref}|^2\right]^{1/2} \right\},
\end{equation}
where
\begin{equation}
\Omega_{\rm ref}\equiv \frac{1}{2}(\gamma_{21}+\Delta\omega_{D}-\gamma_{31}).
\end{equation}
In order to illustrate the quantum interference effect in a simple and clear way, we decompose Im$(K)$
for different $\Omega_a$ as follows.

(i). {\it Weak control field region}  (i.e. $|\Omega_a|<\Omega_{\rm
ref}\approx \Delta\omega_D/2$):  Equation (\ref{Kd1}) can be written as
\begin{equation}\label{Kd2}
K(\omega)=\frac{\omega}{c}+\sqrt{\pi}\kappa_{12}\left(\frac{A_+}{\omega
-\omega_+}+\frac{A_-}{\omega-\omega_-}\right),
\end{equation}
where $A_{\pm}=\mp(\omega_{\pm}+i\gamma_{31})/(\omega_+-\omega_-)$. Since in this
region ${\rm Re}(\omega_{\pm})={\rm Im}(A_{\pm})=0$, we
obtain
\begin{equation}\label{form1}
{\rm
Im}(K)=\sqrt{\pi}\kappa_{12}\left(\frac{B_+}{\omega^2+\delta_+^2}+
\frac{B_-}{\omega^2+\delta_-^2}\right)\equiv L_2+L_1,
\end{equation}
with $\delta_{\pm}={\rm Im}(\omega_{\pm})$, $B_{\pm}=A_{\pm}\delta_{\pm}$, and
$L_{1(2)}=\sqrt{\pi}\kappa_{12}B_{-(+)}/(\omega^2+\delta_{-(+)}^2)$. Thus the probe-field absorption
profile comprises two Lorentzians centered at $\omega=0$.

\begin{figure}
\includegraphics[scale=0.35]{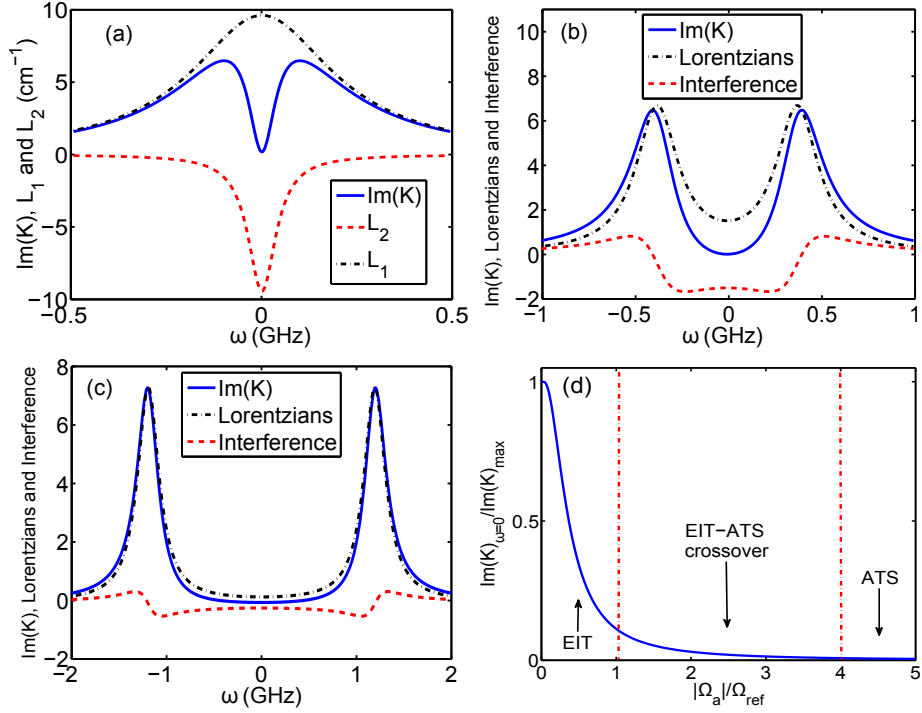}
\caption{\footnotesize (Color online) EIT-ATS crossover for hot
ladder-\uppercase\expandafter{\romannumeral1} system. (a)
Probe-field absorption spectrum Im($K$) (solid line) in the
region $|\Omega_a|<\Omega_{\rm ref}$ is a superposition of the
positive $L_1$ (dash-dotted line) and the negative $L_2$
(dashed line). (b) Im($K$) (solid line) composed by two Lorentzians
(dashed-dotted line) and destructive interference (dashed line)
in the region $|\Omega_a|>\Omega_{\rm ref}$. (c) Im($K$) (solid line)
composed by two Lorentzians (dashed-dotted line) and destructive
interference (dashed line) in the region $|\Omega_a|>\Omega_{\rm ref}$.
Panels (a), (b) and (c) correspond to EIT, EIT-ATS crossover, and ATS,
respectively. (d) The ``phase diagram'' of
${\rm Im}(K)_{\omega=0}/{\rm Im}(K)_{\rm max}$ as a
function of $|\Omega_a|/\Omega_{\rm ref}$ illustrating the transition
from EIT to ATS for the hot ladder-I system. Three regions (EIT,
EIT-ATS crossover, and ATS) are divided by two vertical dashed-dotted lines.}
\label{figEIT1}
\end{figure}

Shown in Fig.~\ref{figEIT1}(a)
%
%
are the results of $L_1$, which is a positive single peak (the
dash-dotted line), and $L_2$, which is a negative single peak (the
dashed line). When plotting the figure, we have taken $\Omega_a=100$ MHz
and the other parameters are the same as those used in Fig.~\ref{fig:LM}(b).
The superposition of $L_1$ and $L_2$ gives Im($K$) (the solid line), which
displays a absorption doublet with a transparency window
near at $\omega=0$. Because there exists a {\it destructive}
interference between the positive $L_1$ and the negative $L_2$
in the probe-field absorption spectrum, the phenomenon found here belongs to
EIT based on the criterion given in
Refs.~\cite{Anisimov2008,Tony2010,Anisimov2011}.

(ii). {\it Intermediate control field region} (i.e. $|\Omega_a|>\Omega_{\rm ref}$):
In this region ${\rm Re}(\omega_{\pm})\neq0$, we obtain
\begin{equation}\label{Kd3}
K(\omega)=\frac{\omega}{c}-\sqrt{\pi}\kappa_{12}\left[\frac{\omega+iW}{(\omega+iW-\delta)
(\omega+iW+\delta)}+\frac{i(\gamma_{31}-W)}{(\omega+iW-\delta)(\omega+iW+\delta)}\right].
\end{equation}
where
$W\equiv(\gamma_{21}+\gamma_{31}+\Delta\omega_D)/2$ and
$\delta\equiv\sqrt{4|\Omega_a|^2-(\gamma_{21}+\Delta\omega_D-\gamma_{31})^2}/2$.
The imaginary part of the  Eq.~(\ref{Kd3}) is given by
\begin{eqnarray}\label{form2}\nonumber
{\rm
Im}(K)=&&\frac{\sqrt{\pi}\kappa_{12}}{2}\left\{\frac{W}{(\omega-\delta)^2+W^2}
+\frac{W}{(\omega+\delta)^2+W^2}\right.\nonumber\\
&&\left.\hspace{0.7cm}+\frac{g}{\delta}\left[\frac{\omega-\delta}{(\omega
-\delta)^2+W^2}-\frac{\omega+\delta}{(\omega+\delta)^2+W^2}\right]\right\},
\end{eqnarray}
with $g=W-\gamma_{31}$. The previous two terms (i.e. the two
Lorentzian terms) in Eq.~(\ref{form2}) can be thought of
as the net contribution coming to the absorption from two different channels
corresponding to the two dressed states created by the control field
$\Omega_a$~\cite{Agarwal1997}. The following terms proportional to $g$
are clearly interference terms. The interference is controlled by the parameter
$g$ and it is destructive (constructive) if $g>0$ ($g<0$).  Since in the
ladder-\uppercase\expandafter{\romannumeral1} system with $x=-1$,
$g=(\gamma_{21}+\Delta\omega_D-\gamma_{31})/2$ is always positive,
thus the quantum interference induced by the control field is always destructive.

Fig.~\ref{figEIT1}(b) shows the probe-field absorption spectrum Im($K$) (solid
line) as a function of $\omega$ for $|\Omega_a|>\Omega_{\rm ref}$. The
dashed-dotted (dashed) line denotes the contribution by the two
positive Lorentzians  (negative interference terms). We see that the
interference is destructive. The system parameters used are the same as those
in Fig.~\ref{figEIT1}(a) but with $\Omega_a=400$ MHz. A transparency window
is opened due to the combined effect of EIT and ATS, which is deeper and wider
than that in Fig.~\ref{figEIT1}(a). We attribute such phenomenon
as EIT-ATS crossover.

(iii). {\it Large control field region} (i.e., $|\Omega_a|\gg\Omega_{\rm ref}$):
In this case, the quantum interference strength $g/\delta$ in Eq.~(\ref{form2})
is very weak (i.e. $g/\delta\approx 0$).  Im$(K)$ reduces to
\begin{equation}\label{form3}
{\rm
Im}(K)=\frac{\sqrt{\pi}\kappa_{12}}{2}\left[\frac{W}{(\omega-\delta)^2
+W^2}+\frac{W}{(\omega+\delta)^2+W^2}\right].
\end{equation}
Fig.~\ref{figEIT1}(c) shows the result of the
probe-field absorption spectrum as a function of $\omega$ for
$|\Omega_a| \gg \Omega_{\rm ref}$. The dashed-dotted line represents
the contribution by the sum of the two Lorentzians. For
illustration, we have also plotted the contribution from the small
interference terms [neglected in Eq.~(\ref{form2})\,], denoted by
the dashed line. We see that the interference is still destructive
but very small. The solid line is the curve of Im($K$), which has
two resonances at $\omega\approx\pm\Omega_a$. Parameters used are
the same as those in Fig.~\ref{figEIT1}(a) and Fig.~\ref{figEIT1}(b)
but with $\Omega_a=1.2$ GHz. Obviously, the phenomenon found in this case
belongs to ATS because the transparency window opened is mainly due to the
contribution of the two Lorentzians.

From the results given above, we see that the probe-field absorption
spectrum experiences a transition from EIT to ATS as the control
field is changed from small to large values. From the above result
we can distinguish three different regions, i.e. the EIT  ($|\Omega_a|<\Omega_{\rm
ref}$), the EIT-ATS crossover ($1< |\Omega_a|/\Omega_{\rm ref}\leq
4$), and ATS $(|\Omega_a|/\Omega_{\rm ref}> 4$).
Fig.~\ref{figEIT1}(d) shows a ``phase diagram'' that illustrates the
transition from the EIT to ATS  by plotting ${\rm
Im}(K)_{\omega=0}/{\rm Im}(K)_{\rm max}$ as a function of
$|\Omega_a|/\Omega_{\rm ref}$. Note that we have defined ${\rm
Im}(K)_{\omega=0}/{\rm Im}(K)_{\rm max}=0.01$ as the border between
EIT-ATS crossover and ATS regions. Our results on the characters of the quantum
interference effect in the hot Rubidium atomic gases are consistent with those obtained
in the experiments~\cite{Julio1995,Moon}. According to our analysis,
the experiments carried out in Refs.~\cite{Julio1995,Moon}
are mainly in the EIT region. We expect the EIT-ATS crossover and ATS may be
observed experimentally if $\Omega_a$ is increased
to the intermediate and the large control-field regions.

\subsection{EIT-ATS crossover in hot molecular gases}\label{sec3c}

In 2008, Lazoudis {\it et al.}~\cite{Lazoudis2008} made an important
experimental observation on EIT and ATS in a hot Na$_2$ molecular ladder-I
system for the wavenumber ratio $x=-0.896$ and
$x=-1.08$~\cite{note1}. Two excitation schemes of Na$_2$ molecules were adopted
in Ref.~\cite{Lazoudis2008}. The first (called the system B)
is $X^1\sum_g^+(1,19)\rightarrow A^1\sum_u^+(3,18)\rightarrow 4^1\sum_g^+(0,17)$, and the
second (called the system A) is $X^1\sum_g^+(0,19)\rightarrow A^1\sum_u^+(0,20)\rightarrow 2^1\Pi_g(0,19)$.
Both of them correspond to the levels $|1\rangle\rightarrow |2\rangle\rightarrow |3\rangle$
in our Fig.~\ref{model}(b). We now analyze this system by using the
Eq.~(\ref{LDa}).

When $x$ is different from  $-1$, the approach used in the last subsection is not
easy to implement since the pole of the integrand in the Eq.~(\ref{LDa}) is not
fixed in the lower (or upper) half complex plane of $v$.
In this case, the value of the pole depends on both $x$ and $\omega$; moreover,
it has an intersection with the real axis for $\omega=0$. As a result, the residue of the pole
is a piecewise function, and the spectrum decomposition gives very complicated expressions
not convenient for analyzing the quantum interference character of the system.

Because of the above mentioned difficulty, we turn to adopt the fitting method
developed from the spectrum decomposition method, proposed by
Anisimov {\it et al.}~\cite{Anisimov2011}.  According
to the spectrum-decomposition formulas~(\ref{form1}) and
(\ref{form3}), we expect: (i)if the probe-field absorption
spectrum has a good fit to the function
\begin{equation}\label{AEIT}
A_{\rm
EIT}=\frac{B_+^2}{\omega^2+\delta_+^2}-\frac{B_-^2}{\omega^2+\delta_-^2},
\end{equation}
EIT dominates, where $B_+,\delta_+,B_-,\delta_-$ are
fitting parameters;
(ii)if the absorption spectrum has a good fit to
the function
\begin{equation}\label{AATS}
A_{\rm
ATS}=C\left[\frac{1}{(\omega-\delta)^2+W^2}+\frac{1}{(\omega+\delta)^2+W^2}\right],
\end{equation}
ATS dominates, with $C,\delta,W$ being fitting parameters.

\begin{figure}
\includegraphics[scale=0.35]{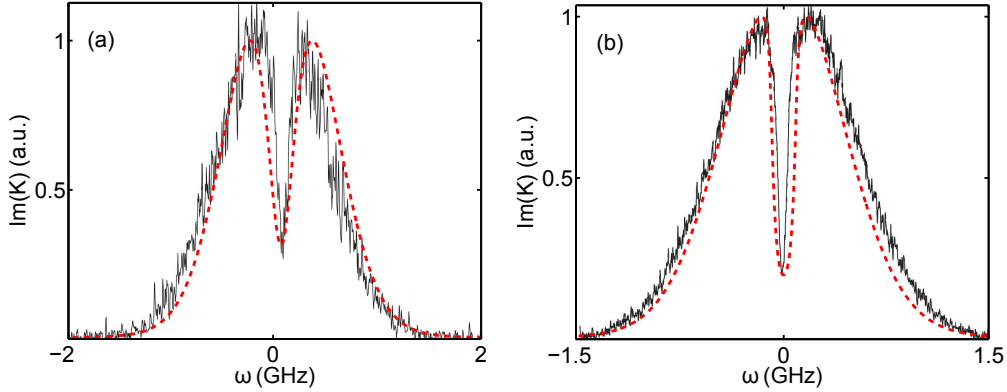}\\
\caption{\footnotesize (Color online) The probe absorption spectrum
Im($K$) as a function of $\omega$ for (a) $x=-0.896$
and $\Omega_a=265$ MHz (corresponding to system B of
Ref.~\cite{Lazoudis2008}), and (b) $x=-1.08$ and $\Omega_a=242.5$
MHz (corresponding to system A of Ref.~\cite{Lazoudis2008}). The red
dashed lines are our theoretical results, and the black-solid lines
are the experimental ones from Ref.~\cite{Lazoudis2008}.} \label{sysA}
\end{figure}

Based on such technique, we find that EIT, ATS, and EIT-ATS crossover exist in the
open molecular ladder-I system for both $x=-1.08$ and $x=-0.896$.
Fig.~\ref{sysA}(a) shows 
%
%
the probe-field absorption spectrum Im($K$) for $x=-0.896$ and $\Omega_a=265$ MHz
(corresponding to system B in Ref.~\cite{Lazoudis2008}). The black
solid line is the experimental result from Ref.~\cite{Lazoudis2008},
while the red dashed line is given by our theoretical
calculation. The system parameters are given by
$\Gamma_{12}=\Gamma_{42}=4.0\times10^{7}$ $\mathrm{s}^{-1}$,
$\Gamma_{23}=5.6\times10^{6}$ $\mathrm{s}^{-1}$,
$\Gamma_{53}=5.0\times10^{7}$ $\mathrm{s}^{-1}$,
$\gamma=2.7\times10^{5}$ $\mathrm{s}^{-1}$, $\gamma_{jl}^{{\rm
col}}=1\times10^{6}$ $\mathrm{s}^{-1}$, and
$\Delta\omega_D=5\times10^{8}$ $\mathrm{s}^{-1}$. We see that our
theoretical result agrees well with the experimental one.
Note that the value of the reference Rabi frequency $\Omega_{\rm ref}$ is a function of
the wavenumber ratio $x$. When $x=-0.896$, one has
$\Omega_{\rm ref}\simeq 400$ MHz. Thus the system
is in the weak control field region and the phenomenon found
belongs to the EIT. Note in passing that here we have plotted the quantity
Im$(K)$ which is proportional to the fluorescence intensity related to state
$|2\rangle$ because $\sigma_{22}\simeq2|\Omega_b|^2{\rm Im}(K)/\Gamma_2$.

Shown in Fig.~\ref{sysA}(b) is the absorption spectrum Im($K$) for $x=-1.08$
and $\Omega_a=242.5$ MHz (corresponding to system A in
Ref.~\cite{Lazoudis2008}). The system parameters are the same as
that in Fig.~\ref{sysA}(a). We see that our result also agrees
well with the experimental one. Since in this case $\Omega_{\rm ref}\approx 150$ MHz,
the system is in the intermediate control
field region and hence the phenomenon found belongs to the EIT-ATS
crossover. Note that there is a small difference for the width of the EIT transparency window
between our result and that in the experiment~\cite{Lazoudis2008}. The reason is mainly
due to the approximation using the modified Lorentzian velocity
distribution to replace the Maxwellian velocity distribution.

\subsection{EIT in hot Rydberg atomic gases}\label{sec3d}

Recently, much interest has focused on the EIT in hot Rydberg atomic gases due to
its promising applications for storing,
manipulating quantum information and precision
spectroscopy~\cite{saf,moh1,moh2,wea,rai,pri,pri1,sevi,Lazoudis2008,Yang1997,Yong1995,Julio1995,
Lee2000,Jason2001,Qi2002,Ahmed2006,Ahmed2007,Ray2007,Moon,Kubler2010,gor,petro,ate}.
The ladder-I system  has been widely adopted in the experimental study of
Rydberg EIT, in which the transition is
$5S_{1/2} \rightarrow  5P_{3/2} \rightarrow  nD_{5/2}$ of $^{85}$Rb atoms with $n$ being a
large integer number.  In this case, the upper state $|3\rangle$ in
Fig.~\ref{model}(c) is a Rydberg state.  If the density
(e.g. lower than $10^8$ cm$^{-3}$) of a Rydberg gas is low, the
interaction between Rydberg atoms can be ignored. Our theory
developed in Sec.~\ref{sec2} and Sec.~\ref{sec3a} can be applied to study the
probe-field propagation in such system.

\begin{figure}
\includegraphics[scale=0.3]{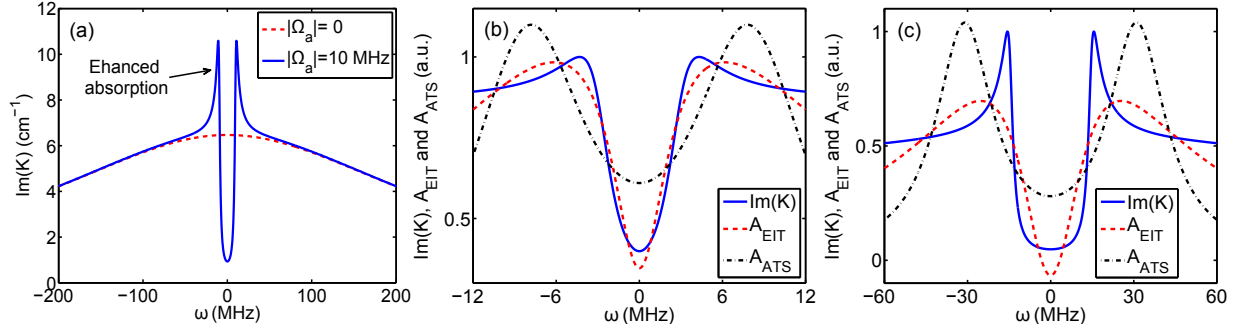}
\caption{\footnotesize (Color online) (a) Probe-field absorption spectrum
Im($K$) as a function of $\omega$. The blue solid (red dashed) line
is for $|\Omega_a|=10$ MHz ($|\Omega_a|=0$). (b)
Im($K$) (blue solid line), $A_{\rm EIT}$
(red dashed line) and $A_{\rm ATS}$ (black dashed-dotted
line) as a function of $\omega$ for the weak control-field $\Omega_a=3$ MHz
where $A_{\rm EIT}$ has a good fit. (c) The case for the intermediate control field
$\Omega_a=15$ MHz where both $A_{\rm EIT}$ and $A_{\rm ATS}$ have poor fit.}
\label{fitEIT}
\end{figure}

Shown in Fig.~\ref{fitEIT}(a)
%
%
is the numerical result of the probe-field absorption spectrum
Im($K$) as a function of $\omega$ for the hot ladder-I system with wavenumber ratio $x=-1.63$,
which corresponds to the experiment carried out in 2007~\cite{moh1} by Mohapatra {\it et al}.
The red dashed (blue solid) line is for the case of $|\Omega_a|=0$ ($|\Omega_a|=10$
MHz) for the system parameters $\Gamma_{2}=6$ MHz, $\Gamma_{3}=1$
kHz, $\gamma_{jl}^{{\rm col}}=1$ MHz, $\Delta\omega_D=270$ MHz, and
$\kappa_{12}=1\times10^{9}$ cm$^{-1}$s$^{-1}$. We find that the line
shape of Im($K$) displays enhanced absorption on both sides of the
transparency window. This effect arises due to the wavelength
mismatch between the control and probe fields combined with the
effect of Doppler broadening. We now analyze the
quantum interference character of such system.

Since the spectrum decomposition method is not convenient for the analysis
for the case $x\neq -1$, we employ the fitting method as done in
the last subsection. Shown in Fig.~\ref{fitEIT}(b) and Fig.~\ref{fitEIT}(c)
are the results of Im($K$) (blue solid line),
$A_{\rm EIT}(B_+,\delta_+,B_-,\delta_-)$ (red dashed line) and $A_{\rm
ATS}(C,\delta,W)$ (black dash-dotted line) as a function of
$\omega$ for $\Omega_a=3$ MHz and 15 MHz, respectively.
The expressions of $A_{\rm EIT}$ and $A_{\rm ATS}$ have been given
by  Eqs.~(\ref{AEIT}) and (\ref{AATS}). From Fig.~\ref{fitEIT}(b)
we see that Im($K$) has a
good fit to $A_{\rm EIT}(3.04,1.58,0.0381,0.208)$ and a poor fit to
$A_{\rm ATS}(0.237,0.686,0.513)$. Thus EIT occurs in this
weak control field region. However, for intermediate and large control field
 one can not find out the fitting parameters
by which $A_{\rm EIT}$ and $A_{\rm ATS}$ can have a good fit to Im($K$)
(Fig.~\ref{fitEIT}(c) shows the result for $\Omega_a=15$ MHz).
Consequently, based on the criterion of Ref.~\cite{Anisimov2011},
neither EIT nor ATS dominates in the intermediate large control field regions.

Note that in the system discussed here the probe-field absorption spectrum
Im($K$) doesn't possess standard Lorentzian lineshape for large control field, which is due to
the enhanced absorption by the Doppler effect  and by the large wavenumber mismatch between
the probe and control fields.
Experimentally, EIT in hot Rydberg atomic gases has been observed in Ref.~\cite{moh1}.
Our theoretical result given above agrees with the experimental one.
We hope that the theoretical result for the intermediate and large control field region predicted
here may be verified experimentally in near future.

\section{Quantum interference character of ladder-II system}\label{sec4}

If the probe field and the control field in the ladder-I system are exchanged,
we obtain the  ladder-II system (Fig.~\ref{model}(b) with 
$\omega_b=\omega_c$, $\omega_a=\omega_p$). In this case, the Maxwell Eq.~(\ref{ME})
under the SVEA is reduced to 
\begin{equation}\label{maxwell2}
i\left(\frac{\partial}{\partial
z}+\frac{1}{c}\frac{\partial}{\partial t}\right)\Omega_a
+\kappa_{23}\int_{-\infty}^{\infty} dv f(v)\sigma_{32}(v)=0,
\end{equation}
with $\kappa_{23}={\cal
N}\omega_a|\boldsymbol{\mu}_{32}|^2/(2\hbar\varepsilon_0 c)$.

\subsection{Linear dispersion relation}

The base state solution of the MB Eqs.~(\ref{dme1}) and (\ref{maxwell2})
of the ladder-II system reads
\begin{subequations} \label{BS}
\begin{eqnarray}
&&\sigma_{11}^{(0)}=\left(\gamma\Gamma_2|d_{21}|^2
+2\gamma\gamma_{21}|\Omega_b|^2\right)\frac{1}{D_1},\\
&&\sigma_{22}^{(0)}=2\gamma\gamma_{21}|\Omega_b|^2\frac{1}{D_1},\\
&&\sigma_{44}^{(0)}=2\gamma_{21}\Gamma_{42}|\Omega_b|^2\frac{1}{D_1},\\
&&\sigma_{21}^{(0)}=-\gamma\Gamma_{2}\Omega_bd_{21}^*\frac{1}{D_1},
\end{eqnarray}
\end{subequations}
and $\sigma^{(0)}_{31}=\sigma^{(0)}_{32}=\sigma^{(0)}_{33}=\sigma^{(0)}_{55}=0$,
with $D_1 \equiv
\gamma\Gamma_2|d_{21}|^2+2\gamma_{21}(2\gamma+\Gamma_{42})|\Omega_b|^2$.

\begin{figure}
\includegraphics[scale=0.4]{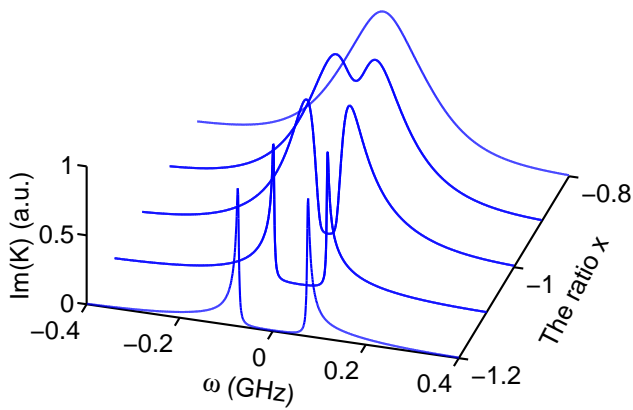}
\caption{\footnotesize (Color online) The probe-field absorption
spectrum Im($K$) of the ladder-II system as a function of $\omega$
and the wavenumber ratio $x=k_a/k_b$.} \label{ATS3D}
\end{figure}

By using the same method as in Sec.~\ref{sec3a}, one can obtain the solution
of the MB Eqs.~(\ref{dme1}) and (\ref{maxwell2})
in linear regime,  with the linear dispersion relation given by
\begin{equation}\label{LDb}
K(\omega)=\frac{\omega}{c}+\kappa_{23}\int_{-\infty}^{\infty} dv
f(v)\frac{(\omega+d_{31})2\gamma\gamma_{21}|\Omega_b|^2-\gamma
\Gamma_{2}|\Omega_b|^2d_{21}^*}{D_1\left[|\Omega_b|^2
-(\omega+d_{31})(\omega+d_{32})\right]}.
\end{equation}
Fig.~\ref{ATS3D} shows
%
%
the probe-field absorption spectrum Im($K$) as a function of $\omega$ and the
wavenumber ratio $x$. We see that, similar to the ladder-I system
(Fig.~\ref{EIT3D}), Im($K$)
undergoes also a transition from a wide transparency window in the line center to
a single absorption peak when $x$ changes from $-1.2$ to
$-0.8$. The system parameters have been chosen as
$\Gamma_2=6$ MHz, $\Gamma_3=1$ MHz, $\gamma=0.5$ MHz,
$\gamma_{ij}^{\rm col}=1$ MHz, and $\Omega_b=100$ MHz.

\subsection{EIT-ATS crossover in hot Sodium atomic gases}

In 1978, Gray and Stroud~\cite{Gray1978} made an experimental
observation on ATS in a ladder-II type hot sodium atomic system 
with
$|1\rangle=|3S_{1/2},F=2,M_F=2\rangle$,
$|2\rangle=|3P_{3/2},F=3,M_F=3\rangle$, 
$|3\rangle=4D_{5/2},F=4,M_F=4\rangle$, 
and the wavenumber ratio $x\approx-1$. 
Such system can be described by
the MB Eqs.~(\ref{dme1}) and  (\ref{maxwell2}), and hence
the linear dispersion relation (\ref{LDb}) can be used to describe
the probe-field propagation.

To get an analytical insight, we replace the Maxwellian
velocity distribution by the modified Lorentzian velocity distribution and calculate the
integration (\ref{LDb}) using the residue theorem. We find two poles of the integrand
in the lower half complex plane of $v$, which are $k_av=-ik_av_T=-i\Delta\omega_D$ and
$k_av=-iC=-i[\gamma_{21}^2+2\gamma_{21}(2\gamma
+\Gamma_{42})|\Omega_b|^2/\gamma\Gamma_{2}]^{1/2}$.  By taking the
contour consisting of the lower half complex plane of $v$ and its real axis, we can
calculate the integration exactly, with the result given by
\begin{equation}\label{Kb1}
K(\omega)=\omega/c+{\cal K}_1+{\cal K}_2,
\end{equation}
with
\begin{subequations} \label{LS}
\begin{eqnarray}
&&{\cal K}_1=\frac{2\sqrt{\pi}\kappa_{23}\gamma_{21}|\Omega_b|^2\left\{\omega
+i[\gamma_{31}+\Gamma_{2}(\Delta\omega_D+\gamma_{21})/(2\gamma_{21})]
\right\}}{\Gamma_{2}(C^2-\Delta\omega_D^2)\left[|\Omega_b|^2-(\omega
+i\gamma_{31})(\omega+i\gamma_{32}+i\Delta\omega_D)\right]},\\
&&{\cal
K}_2=\frac{2\sqrt{\pi}\kappa_{23}\gamma_{21}\Delta\omega_D|\Omega_b|^2
\left\{\omega+i[\gamma_{31}+\Gamma_{2}(C+\gamma_{21})/(2\gamma_{21})]
\right\}}{\Gamma_{2}C(\Delta\omega_D^2-C^2)\left[|\Omega_b|^2
-(\omega+i\gamma_{31})(\omega+i\gamma_{32}+iC)\right]}.
\end{eqnarray}
\end{subequations}

We can also carry out a spectrum decomposition for ${\cal K}_j$ ($j=1,2$),
like that done in Ref.~\ref{sec3a}. The explicit expressions of the decomposition
have been given in Appendix~\ref{AppD}. Similarly,
three different control field regions (i.e. the weak control field region
$|\Omega_b|<\Omega_{\rm ref}$, the intermediate control field region
$|\Omega_b|>\Omega_{\rm ref}$, and the strong control field region
$|\Omega_b|\gg \Omega_{\rm ref}$; $\Omega_{\rm ref}\equiv \Delta\omega_D/2$)
can also be obtained.

\begin{figure}
\includegraphics[scale=0.3]{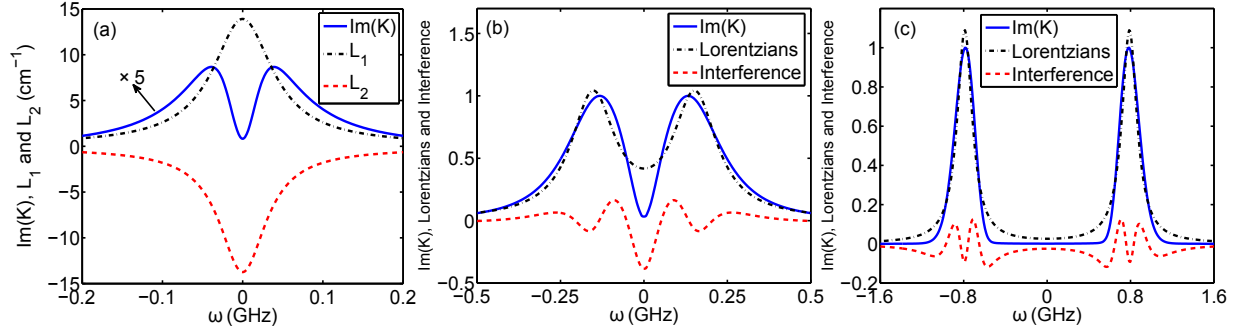}
\caption{\footnotesize (Color online) EIT-ATS crossover for the hot atoms
in the ladder-II system for the wavenumber ratio $x=-1$. (a) Probe-field
absorption spectrum Im($K$) in the weak control field region ($|\Omega_a|<\Omega_{\rm ref}$).
The dashed-dotted line is the contribution by positive $L_1$, the dashed line is by
negative $L_2$. The sum of  $L_1$ and  $L_2$ gives  Im($K$) (solid line).
(b) Probe-field absorption spectrum Im($K$) (solid line) composed by two Lorentzians (dashed-dotted
line) and the destructive interference (dashed line) in the intermediate
control field region $|\Omega_a|>\Omega_{\rm ref}$.
(c) Probe-field absorption spectrum Im($K$) (solid line) composed by two Lorentzians (dashed-dotted
line) and the destructive interference (dashed line) in the strong
control field region  $|\Omega_a|\gg \Omega_{\rm ref}$. Panels (a),
(b) and (c) correspond to EIT, EIT-ATS crossover, and ATS, respectively.} \label{figAT1}
\end{figure}

Fig.~\ref{figAT1}(a)
%
%
shows the absorption spectrum Im($K$) in the weak control field region ($|\Omega_b|=100$ MHz,
which is smaller than $\Omega_{\rm ref}=150$ MHz).
The dashed-dotted line is the contribution by positive $L_1$, and the dashed line is by
negative $L_2$. The superposition (sum) of  $L_1$ and  $L_2$ gives  Im($K$) (solid line). The expressions
of $L_1$ and $L_2$ have been presented in  Appendix~\ref{AppD}.
System parameters are chosen as $\Gamma_2=10$ MHz, $\Gamma_3=3.15$ MHz,
 $\Delta\omega_D=300$ MHz~\cite{Stroud96}, with other parameters the same as those in the last section. We
see that in the curve of Im($K$) a deep transparency window is opened, resulting from
 the {\it destructive} quantum interference (because $L_1$ is positive
and $L_2$ is negative). Hence in this region EIT  exists.

Fig.~\ref{figAT1}(b) shows Im($K$) (solid line)  in the intermediate control field region  
($|\Omega_b|=200$ MHz), which is the sum of the two Lorentzians (dashed-dotted line) and the destructive interference (dashed line). In this region, a large dip appears in Im($K$)
due to the contribution of the destructive
interference. This region belongs to an EIT-ATS crossover.

Fig.~\ref{figAT1}(c) illustrates Im($K$) (solid line),
the two Lorentzians (dashed-dotted line), and the destructive interference
(dashed line)  in the large control field region  ($|\Omega_b|=800$ MHz).
We see that in this region the contribution of the quantum interference is too small
to be neglected. Obviously, the phenomenon found in this situation
belongs to ATS because the transparency window opened is mainly due
to the contribution by the two Lorentzians.

From the above analysis, we see that EIT,
EIT-ATS crossover, and ATS exist in the ladder-II system with the Doppler broadening
for the wavenumber ratio $x=-1$. This is different from cold
ladder-II systems where no EIT and thus EIT-ATS crossover exist~\cite{Tony2010}.
Although the experiment on ATS in a hot atomic system with the ladder-II
configuration for $x=-1$ has been realized~\cite{Gray1978,Stroud96},
it seems that up to now no experimental study has been carried out on EIT,
and EIT-ATS crossover in the ladder-II system with Doppler broadening.
We hope new experiments can be designed to verify our predictions given here.

\subsection{Microwave induced transparency}

We now discuss the case when the control field in the
ladder-\uppercase\expandafter{\romannumeral2} system is a microwave
field, i.e. $x \rightarrow 0$. The relevant experimental result,
named by Zhao {\it et al.}~\cite{Yang1997} as microwave induced transparency,
was first reported  in 1997. 

\begin{figure}
\includegraphics[scale=0.4]{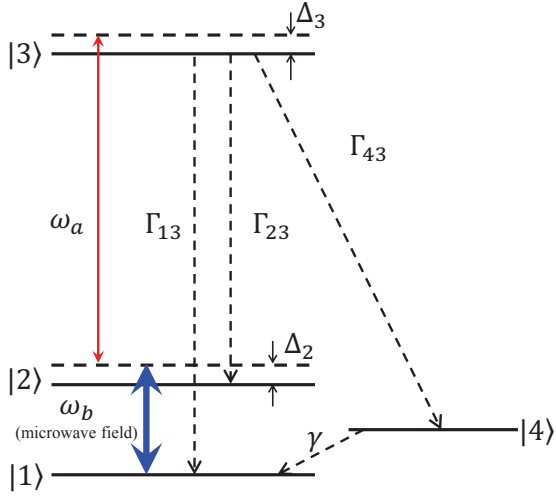}
\caption{\footnotesize (Color online) Microwave field driven
ladder-\uppercase\expandafter{\romannumeral2} configuration.
All notations are given in the text.}
\label{micro}
\end{figure}

In this case, the level diagram and excitation
scheme is given by Fig.~\ref{micro},
%
%
in which  the optical transition between the two lower states $|1\rangle$ and
$|2\rangle$ is forbidden, but the optical transitions between the highest state $|3\rangle$
and the two lower states $|1\rangle$, $|2\rangle$  are allowed,
so $\Gamma_{12}=\Gamma_{42}=0$.  All spontaneous emission decay rates
$\Gamma_{31}$, $\Gamma_{32}$, and $\Gamma_{34}$ (corresponding to the decay pathways
$|3\rangle\rightarrow |1\rangle$, $|3\rangle\rightarrow |2\rangle$, and
$|3\rangle\rightarrow |4\rangle$, respectively), and the transit rate $\gamma$ from
$|4\rangle\rightarrow |3\rangle$ have been indicated in the figure.

The base state solution of the MB equations for the present case
reads $\sigma_{11}^{(0)}=\sigma_{22}^{(0)}=1/2$ and other
$\sigma_{jl}^{(0)}=0$. The linear dispersion relation of the system
is given by
\begin{equation}\label{Kmw}
K(\omega)=\frac{\omega}{c}+\frac{\kappa_{23}}{2}\int_{-\infty}^{\infty}
dv
f(v)\frac{\omega+d_{31}}{|\Omega_b|^2-(\omega+d_{31})(\omega+d_{32})},
\end{equation}
with $d_{31}=-k_av+\Delta_3+i\gamma_{31}$. Because
$\gamma_{31}=\gamma_{32}$, the integrand in Eq.(\ref{Kmw})
has  only one pole in the lower half complex plane of $v$, given by
$k_av=-ik_av_T=-i\Delta\omega_D$.  When replacing the Maxwellian distribution
by the modified Lorentzian distribution, the integration can be calculated
exactly by using the residue theorem. One obtains
\begin{equation}\label{Kmw2}
K(\omega)=\frac{\omega}{c}+\frac{\sqrt{\pi}\kappa_{23}}{2}
\frac{\omega+i\gamma_{31}+i\Delta\omega_D}{|\Omega_b|^2-(\omega
+i\gamma_{31}+i\Delta\omega_D)^2}.
\end{equation}
It is easy to get the probe-field absorption spectrum Im($K$) from Eq.~(\ref{Kmw2}),
which reads
\begin{equation}\label{Kmw3}
{\rm
Im}(K)=\frac{\sqrt{\pi}\kappa_{23}}{2}\left[\frac{W}{(\omega-\delta)^2
+W^2}+\frac{W}{(\omega+\delta)^2+W^2}\right],
\end{equation}
with $W=\gamma_{31}+\Delta\omega_D$ and $\delta=|\Omega_b|$. Equation (\ref{Kmw3})
consists of two pure Lorentzians, which means that there is no quantum interference occurring
in the system and the phenomenon found is an ATS one. Consequently, we conclude that there is no EIT
and EIT-ATS crossover in the ladder-\uppercase\expandafter{\romannumeral2}
system when the control field used is a microwave one.

\section{Summary}\label{sec5}

In Sec.~\ref{sec3} and  Sec.~\ref{sec4}, we have analyzed the quantum interference characters in the hot
ladder-I and ladder-II systems with Doppler broadening for many different
cases. For clearness and for comparison, in Table~\ref{table:SUM}
%
\begin{table}
\caption{Quantum interference characters for various ladder systems with different
wavenumber ratio $x$.  ``Hot'' (``Cold'') means hot (cold) atoms or molecules.
``Any'' means any value of $x$.  The last column gives some references
in which related experiments have been carried out.
\label{table:SUM}}
\begin{ruledtabular}
\begin{tabular}{ccccc}
  System & Wavenumber ratio $x$ & EIT & ATS & Reference \\
  \hline
  \multirow{4}*{Ladder-\uppercase\expandafter{\romannumeral1} (Hot)} & $-0.896$
    & Yes & Yes & \cite{Lazoudis2008} \\
    & $-1$ & Yes & Yes  & \cite{Julio1995,Moon} \\
  & $-1.08$  & Yes & Yes & \cite{Lazoudis2008} \\
  & $-1.63$  & Yes &  No & \cite{moh1}\\\hline
  \multirow{2}*{Ladder-\uppercase\expandafter{\romannumeral2} (Hot)} & $-1$ & Yes & Yes   & \cite{Gray1978} \\
  & 0 & No &  Yes & \cite{Yang1997}\\\hline
  Ladder-\uppercase\expandafter{\romannumeral1} (Cold)& Any   & Yes &
  Yes & \cite{Jason2001,Weatherill2008}
  \\\hline
  Ladder-\uppercase\expandafter{\romannumeral2} (Cold)& Any  & No & Yes  & \cite{Teo2003,Hao2013} \\
\end{tabular}
\end{ruledtabular}
\end{table}
%
we have summarized the main results obtained for different ladder configurations
with different wavenumver ratio $x$. The first four lines are for the hot ladder-I
system; the next two lines are for the hot ladder-II system. The seventh and eighth lines
are for cold ladder-I system and cold ladder-II system, for which relevant theoretical analysis
has been given in Refs.~\cite{Agarwal1997,Tony2010} and related experiments were made in
Refs.~\cite{Jason2001,Weatherill2008,Teo2003,Hao2013}. If in the table there is ``Yes''
in the same line for both EIT and ATS, an EIT-ATS crossover also exists
in the system. The last column of the table gives some references in which related
experimental results were reported.

In summary, in this work we have proposed a general theoretical scheme for studying 
the crossover from EIT to ATS in the open systems of ladder-type level
configuration with Doppler broadening. We have elucidated various mechanisms of
the EIT, ATS, and their crossover in such systems in a clear and unified way.
We have obtained the following conclusions.
First, when the wavenumber ratio $x\approx -1$, EIT, ATS, and EIT-ATS crossover
exist for both ladder-I and ladder-II systems.  Second, when $x$ is far from $-1$,
EIT can occur but ATS is destroyed if the upper state of the ladder-I system is a
Rydberg state. Third, ATS exists but EIT is not possible if the control field that couples the two lower states of the ladder-II system is a microwave field.
Our theoretical analysis have applied to various ladder systems (including
hot gases of Rubidium atoms, molecules, and Rydberg atoms, and so on), and the results
obtained on the quantum interference characters agree well with experimental ones
reported up to now. The results obtained here may have practical applications
in optical information processing and transmission.

\begin{acknowledgments}

This work was supported by NSF-China under Grant Nos. 10874043 and 11174080.

\end{acknowledgments}

\appendix
\section{Spectrum decomposition of the ladder-II system for the wavenumber ratio $x=-1$}\label{AppD}

${\cal K}_j$ $(j=1,2)$ in Eq.~(\ref{LS}) can be decomposed as the form
\begin{equation}\label{decom1}
{\cal K}_j=\eta_j \left(
\frac{A_{j+}}{\omega-\delta_{j+}}+\frac{A_{j-}}{\omega-\delta_{j-}}
\right),
\end{equation}
where $\eta_j$, $A_{j\pm}$ are constants, $\delta_{j+}$ and
$\delta_{j-}$ are two spectrum poles of ${\cal K}_j$, given by
\begin{subequations}\label{corpoles}
\begin{eqnarray}
&&\eta_1=\frac{2\sqrt{\pi}\kappa_{23}\gamma_{21}|\Omega_b|^2}{\Gamma_2(C^2-\Delta\omega_D^2)},\\
&&\eta_2=\frac{2\sqrt{\pi}\kappa_{23}\gamma_{21}\Delta\omega_D|\Omega_b|^2}{\Gamma_2C(\Delta\omega_D^2-C^2)},\\
&&\delta_{1\pm}=\frac{1}{2}\left[-i(\gamma_{32}+\Delta\omega_D+\gamma_{31})\pm\sqrt{4|\Omega_b|^2-(\gamma_{32}+\Delta\omega_D-\gamma_{31})^2}\right],\\
&&\delta_{2\pm}=\frac{1}{2}\left[-i(\gamma_{32}+C+\gamma_{31})\pm\sqrt{4|\Omega_b|^2-(\gamma_{32}+C-\gamma_{31})^2}\right],\\
&&A_{1\pm}=\mp\left\{\delta_{1\pm}-\left[\gamma_{31}+\frac{\Gamma_2}{2\gamma_{21}}(\Delta\omega_D+\gamma_{21})\right]\right\}/(\delta_{1+}-\delta_{1-}),\\
&&A_{2\pm}=\mp\left\{\delta_{2\pm}-\left[\gamma_{31}+\frac{\Gamma_2}{2\gamma_{21}}(C+\gamma_{21})\right]\right\}/(\delta_{2+}-\delta_{2-}).
\end{eqnarray}
\end{subequations}
In order to illustrate the quantum interference effect in a simple
and clear way, we decompose Im$({\cal K}_j)$ in different control field regions as follows.

(i).{\it Weak control field region}  (i.e. $|\Omega_b|<\Omega_{\rm ref}\approx \Delta\omega_D/2$):
In this region, one has Re$(\delta_{j\pm})$=0,
Im$(A_{j\pm})$=0,  and hence
\begin{equation}\label{A_weak}
{\rm Im}(K)=\sum_{j=1}^{2}{\rm Im}({\cal
K}_{j})=\sum_{j=1}^{2}\eta_{j}\left(\frac{B_{j+}}{\omega^2+W_{j+}^2}
+\frac{B_{j-}}{\omega^2+W_{j-}^2}\right)=L_1+L_2,
\end{equation}
where $L_1$ and $L_2$ are defined by
\begin{subequations}\label{L}
\begin{eqnarray}
&&L_1=\frac{\eta_1B_{1-}}{\omega^2+W_{1-}^2}+\frac{\eta_2B_{2-}}{\omega^2+W_{2-}^2}\\
&&L_2=\frac{\eta_1B_{1+}}{\omega^2+W_{1+}^2}+\frac{\eta_2B_{2+}}{\omega^2+W_{2+}^2},
\end{eqnarray}
\end{subequations}
with the real constants
\begin{subequations}\label{Wpm}
\begin{eqnarray}
& & C_{j+}=-W_{j+}(W_{j+}+\Gamma^w_j)/(W_{j+}-W_{j-}),\\
& & C_{j-}=W_{j-}(W_{j-}+\Gamma^w_j)/(W_{j+}-W_{j-}),\\
& &
W_{1\pm}=\frac{1}{2}\left[\gamma_{32}+\gamma_{31}+\Delta\omega_D\pm\sqrt{[\gamma_{32}
+\Delta\omega_D-\gamma_{31}]^2-4|\Omega_b|^2}\right],\\
& &
W_{2\pm}=\frac{1}{2}\left[\gamma_{32}+\gamma_{31}+C\pm\sqrt{[\gamma_{32}
+C-\gamma_{31}]^2-4|\Omega_b|^2}\right],\\
&&\Gamma_1^w=\gamma_{31}+\frac{\Gamma_2}{2\gamma_{21}}(\Delta\omega_D+\gamma_{21}),\\
&&\Gamma_2^w=\gamma_{31}+\frac{\Gamma_2}{2\gamma_{21}}(C+\gamma_{21}).
\end{eqnarray}
\end{subequations}

(ii).{\it Intermediate control field region} (i.e. $|\Omega_b|>\Omega_{\rm
ref}$): By extending the approach by Agarwal~\cite{Agarwal1997}, we
can decompose Im$({\cal K}_j)$ ($j=1,2$) as the form
\begin{eqnarray}\label{A_inter}
{\rm Im}({\cal
K}_{j})=\eta_{j}&&\left\{\frac{1}{2}\left[\frac{W_{j}}{(\omega-\delta_{j}^r)^2
+W_{j}^2}+\frac{W_{j}}{(\omega+\delta_{j}^r)^2+W_{j}^2}\right]\right.\nonumber\\
&
&\left.+\frac{g_{j}}{2\delta_{j}^r}\left[\frac{\omega-\delta_{j}^r}{(\omega
-\delta_{j}^r)^2+W_{j}^2}-\frac{\omega+\delta_{j}^r}{(\omega+\delta_{j}^r)^2+W_{j}^2}\right]\right\},
\end{eqnarray}
where
\begin{subequations}\label{CS1co}
\begin{eqnarray}
&&W_1=(\gamma_{31}+\gamma_{32}+\Delta\omega_D)/2,\\
&&W_2=(\gamma_{31}+\gamma_{32}+C)/2,\\
&&\delta^r_{1}=\sqrt{4|\Omega_b|^2-(\gamma_{32}+\Delta\omega_D-\gamma_{31})^2}/2,\\
&&\delta^r_{2}=\sqrt{4|\Omega_b|^2-(\gamma_{32}+C-\gamma_{31})^2}/2,\\
&&g_1=-\frac{\Gamma_{2}}{4}+\frac{\gamma_{21}-\Gamma_{2}}{2\gamma_{21}}\Delta\omega_D,\\
&&g_2=-\frac{\Gamma_{2}}{4}+\frac{\gamma_{21}-\Gamma_{2}}{2\gamma_{21}}C.
\end{eqnarray}
\end{subequations}

(iii).{\it Large control field region} (i.e.
$|\Omega_b|\gg\Omega_{\rm ref}$): In this case, the quantum
interference strength $g_j/\delta_j^r$ in Eq.~(\ref{A_inter}) is
very weak and negligible. We have
\begin{equation}\label{A_strong}
{\rm Im}({\cal K}_j)\approx
\frac{\eta_j}{2}\left[\frac{W_j}{(\omega-\delta_j^r)^2+W_j^2}
+\frac{W_j}{(\omega+\delta_j^r)^2+W_j^2}\right].
\end{equation}
%


\end{document}